\documentclass[prl,twocolumn,showpacs,preprintnumbers,amsmath,amssymb,superscriptaddress]{revtex4}
\usepackage{graphicx}

\begin{document}

\title{Dynamical vertex approximation for nanoscopic systems}
\author{A.~Valli$^1$, G.~Sangiovanni$^1$, O.~Gunnarsson$^2$, A.~Toschi$^1$, and K.~Held}
\affiliation{ Institute of Solid State Physics, Vienna University of Technology, 1040 Vienna, Austria\\
$^2$ Max-Planck-Institut f\"ur Festk\"orperforschung, 70569 Stuttgart, 
Germany}
\date{\today}

\begin{abstract}
With an increasing complexity of 
nanoscopic systems and the modeling thereof, new theoretical tools are needed for
a reliable calculation of complex systems with strong
electronic correlations. To this end, we 
propose a new approach based on the recently introduced dynamical vertex
approximation. We demonstrate its reliability already on the one-particle vertex (i.e., dynamical mean field theory) level by comparison
with the exact solution. Modeling a quantum point contact
with  110  atoms,  we show that the contact becomes insulating
already before entering the tunneling regime due to a local Mott-Hubbard
transition occurring on the atoms which form the point contact. 
\end{abstract}

\maketitle

\let\n=\nu \let\o =\omega \let\s=\sigma


{\em Introduction.}
In recent years, we have seen
a tremendous experimental progress in the direction of 
man-made nanostructures.
For example, with the wizardry of modern semiconductor technology
quantum effects in quantum dots could be revealed \cite{Beenakker91a,GoldhaberGordon98a,Cronenwett98a,Schmid98a,Kouwenhoven01a,Wiel03a,Hubel08a}; in the area of molecular
electronics transport through single molecules  can now be studied  \cite{Gimzewski87a,Joachim00a,Hettler03a}; and
for magnetic storage technology nanoclusters of transition metals on
surfaces become relevant \cite{Billas93a}.
In all three examples electronic correlations play a decisive role 
since the restriction to nanostructures
brings the electrons close to each other so that their
mutual Coulomb interaction becomes large (compared to
their kinetic energy or tunneling rates). 
As a matter of fact, electronic correlations are not only genuine
to nanostructures, but they also 
 make them fascinating, both from the basic 
research point of view, with new physics occurring, and
from the point of applications since strong correlations
result in spectacular physical properties. An example is the Kondoesque physics
which overcomes  the Coulomb blockade and which has been
observed in the conductance of quantum dots \cite{GoldhaberGordon98a,Cronenwett98a,Schmid98a, Kouwenhoven01a,Hubel08a}
as well as for  small clusters and individual adatoms  \cite{Madhavan98a,Li98a,Ternes09a,Costi09a}.

The theoretical modeling of strong correlations 
in nanostructures attached to some environment (bath)
such as the source and drain electrode in case of the quantum dot
or the surface for the transition metal cluster
is hitherto based on generalizations of the Anderson impurity model
\cite{Anderson84a,Hewson,Kouwenhoven01a,NRG,Hubel08a}. 
However, if one is not only dealing with a single or two "sites'' (say the 
number of quantum dots), 
the numerical effort to solve the corresponding Anderson impurity model becomes prohibitively expensive.
More precisely, the effort grows exponentially with the number of ``sites''
for an exact \cite{Caffarel,Sangiovanni} or numerical renormalization group \cite{NRG} treatment.
This restricts these methods in effect to ${{\cal O}(2)}$  sites coupled to a bath.
Related dynamical matrix renormalization group (DMRG) approaches \cite{Schollwoeck05a} might allow for slightly larger systems  but ultimately suffer
from the same non-polynomial problem,
except for truly one dimensional geometries.
Potentially more efficient  quantum 
Monte-Carlo methods \cite{Maier04} on the other hand 
exhibit a growing sign problem with increasing system size.
Hence a good theory for correlated nano-systems with even 
a few coupled nano-objects is presently missing.
Such a theory is however mandatory since future technological
applications will require the engineering of complex networks
of such nano-objects
-- be it for a quantum computer or 
for a von Neuman computer based on 
such small structures so that quantum effects
are no longer negligible.

On the other side, 
dynamical mean field theory (DMFT) \cite{Metzner89a,Georges92a,Georges96a,Kotliar04a} along with its
cluster \cite{Maier04} and  diagrammatic extensions such
as the dynamical vertex approximation (D$\Gamma$A)\cite{Toschi07a}
 and the dual fermion approach\cite{DualFermion} has been  applied  to strongly correlated electron systems 
with great success: On the model level, among others, 
the   Mott-Hubbard transition \cite{Georges92a,Georges96a},
magnetism \cite{Jarrell92a},
and kinks in strongly correlated systems \cite{Byczuk06a,Macridin07,Toschi09a}
could be better understood or have even been discovered.
Merging DMFT with density functional theory in
the local density approximation (LDA) \cite{LDADMFT1,LDA++,Kotliar06,Held07a} turned out to be a breakthrough for the calculation
of actual materials with strong correlations.
By construction, these DMFT calculations are done in the thermodynamic limit, i.e., for a
macroscopically large system.

There has been one attempt by S.\ Florens  \cite{Florens08a}
to establish a nanoscopic version of DMFT. The idea behind 
Florens' nanoDMFT approach
is the DMFTesque limit of a large number of neighbors
with a central site in the middle,
surrounded by many neighbors
 which in turn are coupled to many neighbors etc. In such a geometry 
(e.g., of a Cayley-type tree), one gets a recursive method 
where the inner sites depends on their outer neighbors but not vice versa. 
Experimentally however,
such a geometry with more and more neighbors is hardly realizable,
and the approach has been
scarcely  used in practice \cite{footnote1}.

Here,
we  hence take another route based on the 
D$\Gamma$A concept of the locality of the fully irreducible
$n$ particle vertex $\Gamma$. While the  calculations in this first paper
will be for $n\!=\!1$, i.e. on a DMFTish level, we  call the approach
nanoD$\Gamma$A in following -- also to distinguish it
from the aforementioned nanoDMFT.
Below, we introduce the approach, validate
its range of applicability against the exact numerical solution
for system sizes where this is still possible,
and demonstrate its potential by hands of calculations for
a quantum point contact with 110 sites.

\begin{figure}[tb]
\includegraphics[width=7cm]{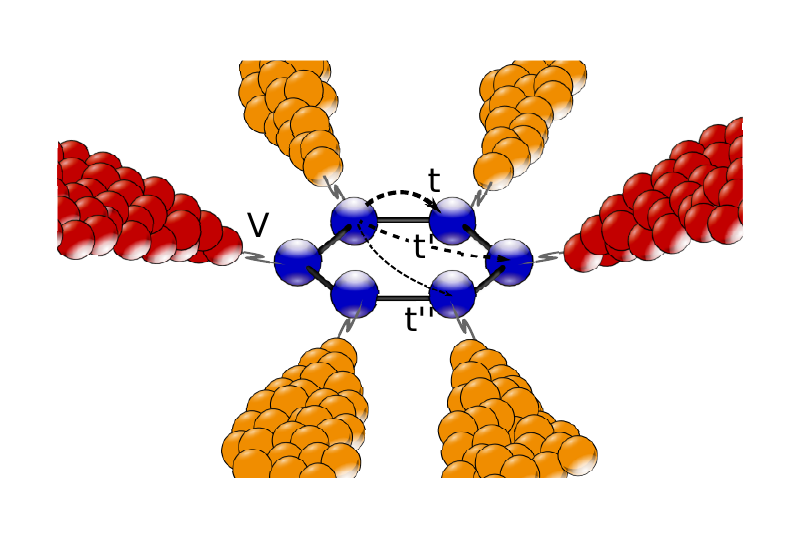}
\includegraphics[width=6.5cm]{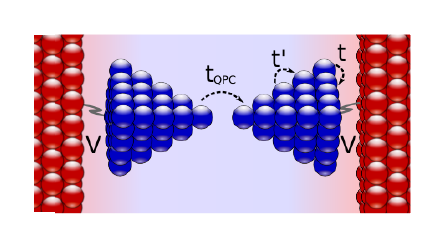}
\caption{Abstract scheme of the nano-systems investigated.
The individual nano-objects (blue sites) are connected by tunneling channels
with amplitude $t$. These  amplitude might be rather complex and
dependent on the distance as indicated by the $t'$ and $t''$ tunneling channels
exemplarily displayed. The whole nano system is
connected via some
 additional tunneling channels $V$ to the environment
sketched schematically in red/orange and assumed to be non-interacting. Possible experimental realizations are coupled
quantum dots or molecules 
connected to source and drain electrodes or a cluster of Fe atoms with the
environment being a surface instead of the indicated  geometry.
Electronic correlations are particularly strong because of the
nanostructuring.
Lower panel: Geometry of the quantum point contact with 110 atoms studied below.
\label{Fig:NanoDGA}
 }
\end{figure}

{\it Method.} As pointed out in the introduction, we are interested in a nanoscopic system consisting of nano-objects (sites) $i$ which are hybridized via $t_{ij}$,
interacting by a Coulomb repulsion $U_{i}$ and coupled by $V_{i\nu k}$ 
to some non-interacting environment, see Fig.\ \ref{Fig:NanoDGA}.
The Hamiltonian hence reads 
\begin{eqnarray}
H&=& \sum_{ij\sigma} t_{ij} c^{\dagger}_{i\sigma} c^{\phantom{\dagger}}_{j\sigma}
+ \sum_i U_{i}  c^{\dagger}_{i\uparrow} c^{\phantom{\dagger}}_{i\uparrow} c^{\dagger}_{i\downarrow} c^{\phantom{\dagger}}_{i\downarrow} \nonumber \\ &&+
 \sum_{i \nu k \sigma} V_{i  \nu k}  c^{\dagger}_{i\sigma}  l^{\phantom{\dagger}}_{\nu k\sigma} + h.c. + \sum_{\nu k \sigma} \epsilon_{\nu k}  l^{\dagger}_{\nu k \sigma}  l^{\phantom{\dagger}}_{\nu k\sigma},
\label{Eq:Hamiltonian}
\end{eqnarray}
where  $c^{\dagger}_{i\sigma}$ ($c^{\phantom{\dagger}}_{i\sigma}$)
and  $l^{\dagger}_{\nu k\uparrow}$ ($l^{\phantom{\dagger}}_{\nu k\uparrow}$)
denote the creation (annihilation) operators for an electron with spin $\sigma$  on site $i$ and in lead $\nu$ state $k$ with energy $\epsilon_{\nu k}$, respectively.  While we consider a single band situation in the following,
Hamiltonian (\ref{Eq:Hamiltonian}) can easily be generalized to include
orbital realism, leading to an additional orbital index in the second quantization operators and orbital matrices in the Green functions and self energies below.

\begin{figure}
\begin{center}
\includegraphics[width=4.5cm,angle=270,clip=true]{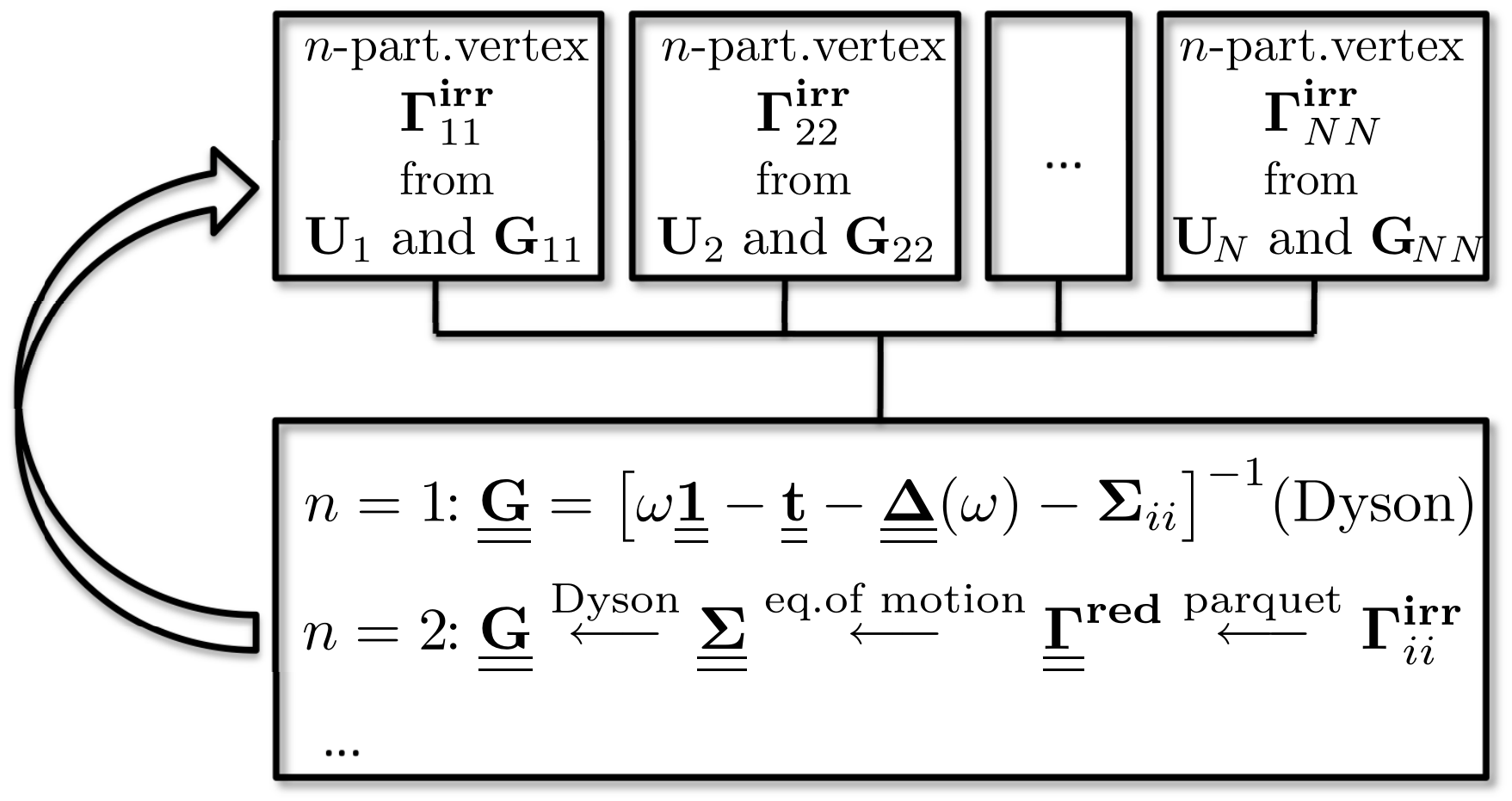}
\end{center}
\caption{NanoD$\Gamma$A algorithm consisting of two self-consistency steps: 1) From the local Green function $G_{ii}$ and Coulomb interaction $U_i$ the $n$-particle fully irreducible vertex $\Gamma^{\rm irr}_{ii}$
has to be calculated. 2) From  $\Gamma^{\rm irr}_{ii}$ a new Green
function $G_{ii}$ is determined through the Dyson and parquet equation, respectively.
\label{Fig:NanoScheme}
 }
\end{figure}

As an exact solution of Hamiltonian (\ref{Eq:Hamiltonian}) is
possible at most for ${\cal O}(10)$ interacting sites, we here propose
an approximate D$\Gamma$A solution.
To this end, we first need to calculate the fully irreducible $n$-particle
vertex on every site $i$ with the interacting Green function
$G_{ii}(\omega)$ and Coulomb interaction $U_{i}$ as an input, see Fig.\ \ref{Fig:NanoScheme}.
In practice, this is done by  numerically calculating the 
corresponding $n$-particle
vertex of an associated Anderson impurity model.
Note that the effort for this computationally most expensive step
only grows {\em linearly} with the number of sites and is easily
parallelizable. 
From the $n$-particle vertex in turn, we recalculate the
Green function and proceed with the first step until convergence.
In the case  $n\!=\!1$, the one-particle fully irreducible  
vertex is simply the self energy $\Sigma(\omega)$ which is
directly related to $G_{ii}(\omega)$ through the
Dyson equation given in Fig.\ \ref{Fig:NanoScheme} in matrix notation
for the site indices $ij$.
For $n\!=\!2$, one needs instead to use the parquet equations 
to go from the irreducible vertex to
 the reducible  one and the exact equation of motion to
get the self energy before proceeding with the Dyson eq., similarly
as discussed in Ref.\onlinecite{Toschi07a} for an infinite system.

Let us note that the approach becomes exact in several limits:
(i) $U\!\rightarrow\! 0$,  (ii) $V \!\rightarrow \!\infty$, 
(iii) number of connections to neighbor sites $Z\!\rightarrow \!\infty$, and, if each site couples to its own lead (iv) $t\!\rightarrow\! 0$. 
While the exact quantum Monte Carlo simulation is impossible for large clusters due to the so-called sign problem, our method is sign problem-free. For $n\!=\!2$ the approach also takes into account the Cooperon diagrams so that weak localization physics is explicitly included, as are spin fluctuations.

{\em Validation vs.\ exact result.}
As a first test case and to validate the approach for $n\!=\!1$
against the numerically exact solution, we consider the
6-site benzene geometry of   Fig.\ \ref{Fig:NanoDGA} (upper panel)
with a constant density of states $\rho$  in the contacts
from $-D$ to $D$ around the Fermi level ($D\! = \!2t$, $t=1$ sets our 
unit of energy in the following); and 
a site-diagonal hybridization $V_{i\nu k}\!=\!V \delta_{i \nu}$.
Two topologies are considered: (i) hopping
$t_{ij}$ restricted to the two nearest neighbors in the hexagon ``(nn $t$)'' and
(ii) an equal hopping to all sites ``(all $t$)''.
Both D$\Gamma$A and the exact solution
are calculated by means of Monte-Carlo simulations \cite{Hirsch86a}
for $U\!=\!5t$,  temperature $T\!=\!0.05t$.

\begin{figure}[t!]
\begin{center}
\includegraphics[width=8cm]{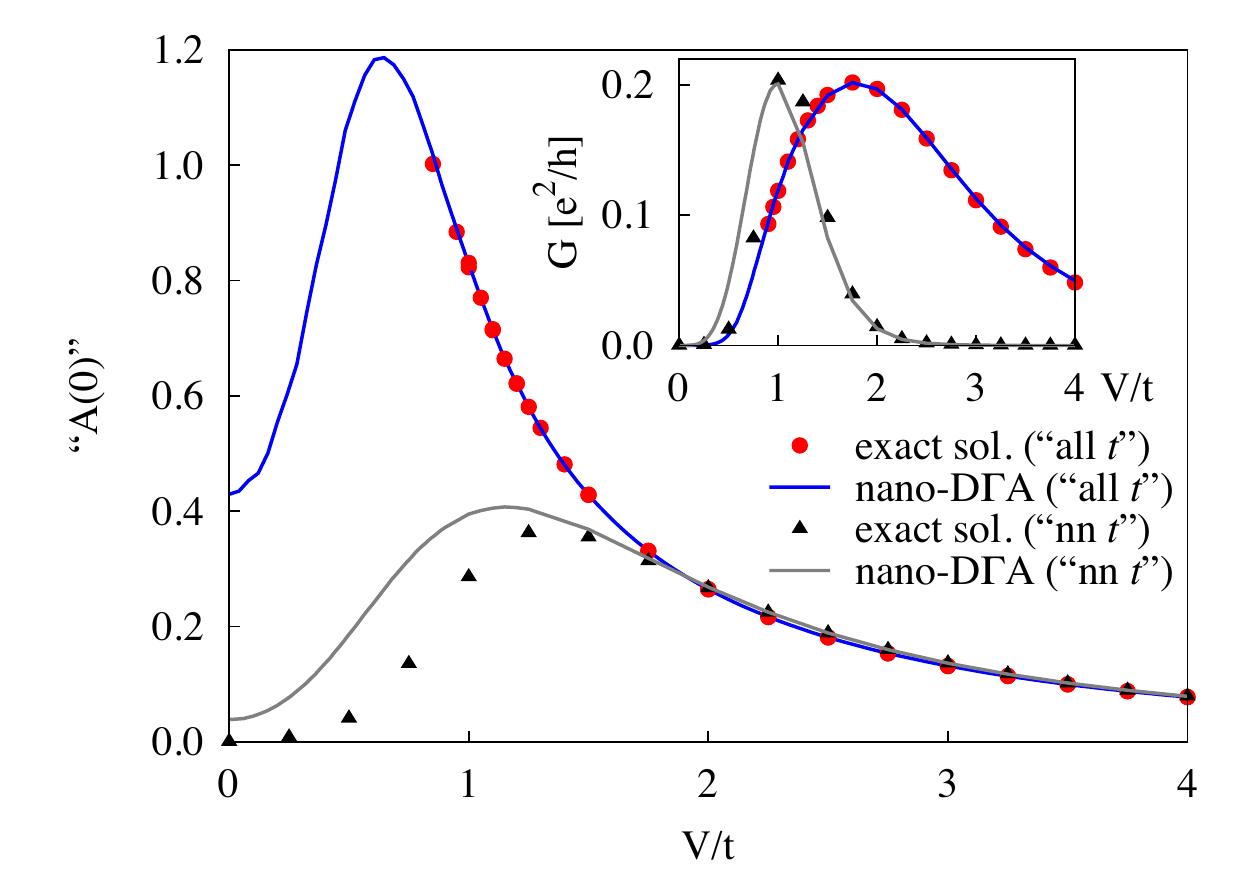}
\end{center}
\caption{(Color onilne)  On-site spectral function $``A(0)"$ 
vs.\ hybridization $V$, comparing
 nanoD$\Gamma$A (lines) with the exact solution (symbols) for two different
geometries: hopping to two nearest neighbors only  and hopping
to all neighbors.
Inset: Conductance between two opposite leads of the benzene ring
i.e., between the dark red leads of Fig. \ref{Fig:NanoDGA},
for both topologies of neighborhood.
Together with the low off-site-diagonal self energy,
the results show that the nanoD$\Gamma$A approach is reliable
if either the hybridization strength is sufficiently strong
or if the inter-cluster hopping is to sufficiently many neighbors.
\label{Fig:benzene}
 }
\end{figure}

Fig.\ \ref{Fig:benzene}  shows the calculated local spectral
function in an interval $2T$ around the Fermi level (set to zero)
$``A(0)"\!=\!\int {\rm d}\omega A(\omega)/[e^{\omega/2T}+e^{-\omega/2T}]$. The results  clearly show that
nanoD$\Gamma$A is reliable both for a hybridization $V\gtrsim t$
and if enough neighbors (in our case 5 in the ``all $t$'' topology) are involved in the hopping
processes. There are 
some deviations for the nearest-neighbor-hopping-only case 
if $V\lesssim t$  since in this situation non-local correlations
such as those involved in forming a two-site singlet are relevant.
If the hopping is to all neighbors though, no deviations
could be identified all the way down to the last calculable point $V\!=\!0.8t$. 
Below $V\!=\!0.8t$, the sign problem becomes too severe
[average sign ${\cal O}(10^{-3})$], and only the nanoD$\Gamma$A
solution is possible - in a situation very favorable for
nanoD$\Gamma$A because of the many neighbors.
Let us note that another indication for the small
correction is  the off-diagonal self-energy, which is smaller than
$10^{-2}t$ for $V>t$ and nearest neighbor hopping as well as
for all $V$ and hopping to all neighbors.

In the inset to Fig.\ \ref{Fig:benzene}, we present the conductance $G$
through  the benzene-like nanostructure, depicted in the upper panel of Fig.\ \ref{Fig:NanoDGA},
from one side of the molecule
to the opposite side, calculated along the lines
of Ref.\onlinecite{Georges99a}.  As 
the local spectral function, the conductance again shows the reliability 
of the calculation already on the one-particle vertex level, with discernible deviations from the exact solution only for 
 a low number of neighbors (i.e., 2 ``nn only'') 
and small hybridization to the leads. 
The results can be understood as follows:
At small hybridization $V$, we have a conductance through two tunneling lead-benzene contacts, leading to an increase of the conductance  $\sim V^4$ (i.e., squared tunneling rate $\Gamma\!=\!\pi V^2 \rho$). 
In the large $V$ region on the other hand, a Kondo resonance 
 between the individual sites of the benzene
molecule and the respective lead forms,
which suppresses the competing  inter-benzene hopping and hence the
transport through the molecule. If hopping between all benzene sites
is allowed (``all $t$''), this effect is less pronounced since there is a direct
hopping channel between the opposite sites to which the
voltage has been applied.

\begin{figure}[t!]
\begin{center}
\includegraphics[width=8cm]{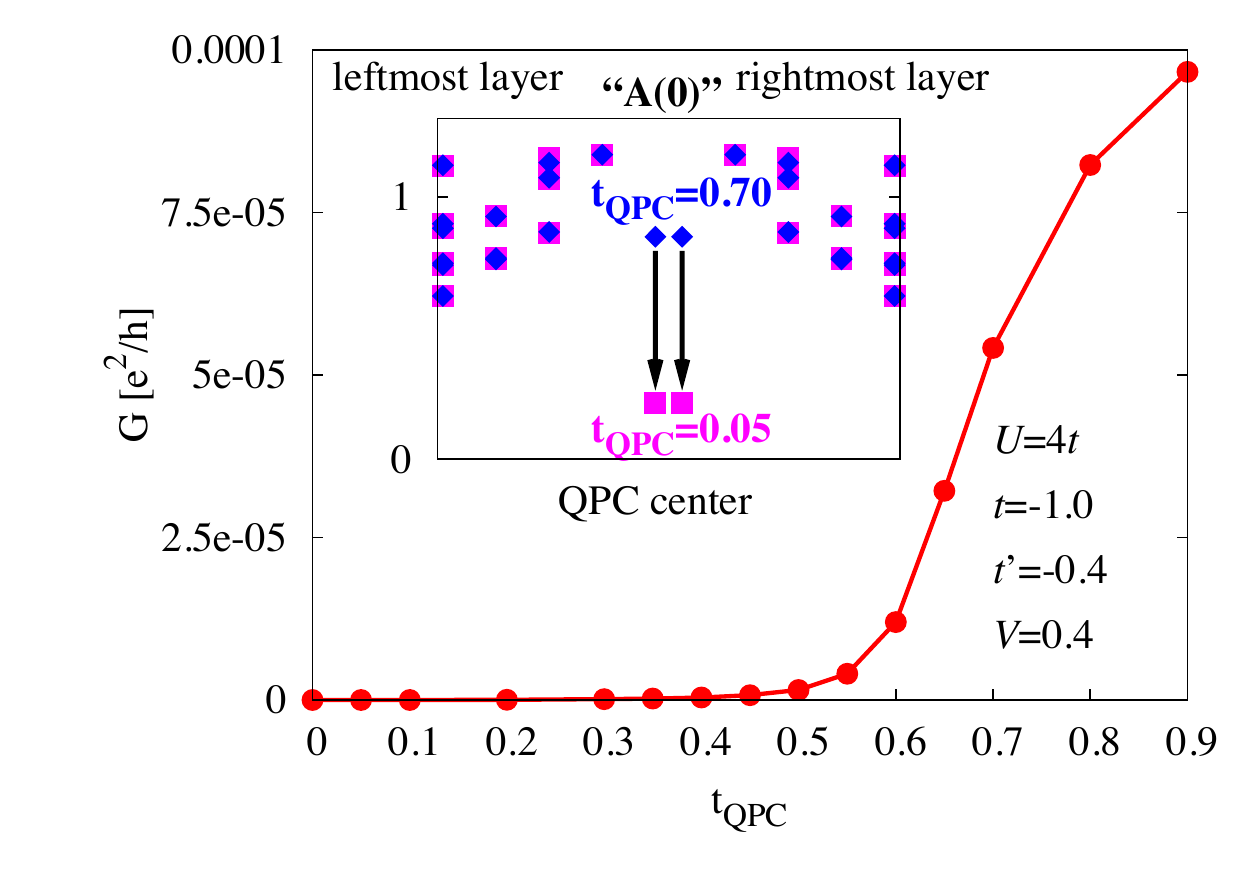}
\end{center}
\caption{(Color online) 
Conductance across the quantum point contact of Fig.\ 
\ref{Fig:NanoDGA} (lower panel) vs. hybridization  $t_{QCP}$ between the
two atoms of the quantum point contact.
The conductance becomes almost zero for $|t_{QCP}|\! \lesssim \!0.5t$.
Inset: Spectral function layer-by-layer across the quantum point contact for two $t_{QCP}$'s below and above the conductance 
increase, revealing a local Mott-Hubbard ``transition'' in the two
atoms forming the QPC.
\label{Fig:QPCconductance}
 }
\end{figure}

{\em Quantum point contact (QPC).}
To demonstrate the suitability of the approach for more complex 
nanosystems, let us now consider a QPC.
Experimentally it can be realized  e.g. through a mechanically 
controllable break junction of a conducting wire, see e.g. Ref.\ \onlinecite{Scheer98a,Agrait03a}.
The assumed geometry is based on a body-centered cubic basic structure narrowed to a double-cone-like junctions as shown in Fig\ \ref{Fig:NanoDGA} (lower panel).
For the moment we assume a single band which might be 
realizable in more complex wires such as cuprate and cobaltate wires, but
realistic calculations, e.g.,  in the spirit of LDA+DMFT \cite{LDADMFT1,LDA++,Kotliar06,Held07a}, are certainly possible. The parameters are: $U\!=\!4t$, $t\!=\!-1$, $t'\!=\!-0.4$, $V\!=\!0.4$, and $T=0.05t$. 
Each calculation takes about 10 hours with a mildly parallelization on 25 Nehalem Intel processors (X5550, 2.66GHz), showing that
much bigger calculations or calculations with orbital realism are possible.
When slowly breaking up the junction 
 the hybridization (tunneling) $t_{\rm QPC}$ between the two atoms forming the point contact will change most strongly and is hence varied.

Surprisingly, we observe a dramatic reduction of the conductance for $|t_{\rm QPC}|\! \lesssim\! 0.5 t$ in Fig.\ \ref{Fig:QPCconductance}. 
Breaking up the junction triggers
a local Mott-Hubbard
``transition'' (more precisely a crossover) of the two
 atoms forming the quantum point,  see   Fig.\ \ref{Fig:QPCconductance} inset. Therefore the conductance drop
with increasing distance between the two atoms is
faster than the exponential decay of $t_{\rm QCP}$ with distance.
Our findings  might explain similar experimental observations
\cite{Krans93a} in transition metals point contacts
 with partially filled $d$-shells,
where the electrons are actually similarly strongly correlated as in our 
calculation. 
This effect could not have been revealed 
with hitherto employed methods \cite{Agrait03a} such as
LDA, Landauer formalism or Coulomb blockade calculations.

{\em Acknowledegment} We acknowledge financial support from the EU network MONAMI and the FWF through the ``Lise-Meitner'' grant n.M1136 (GS) and science college WK004. We thank A.~Georges, P.~Hansmann, D.~Jacob, A.~Katanin, S.~Lichtenstein, and S.~Okamoto for discussions as well as the KITP Santa Barbara for hospitality.


\begin{thebibliography}{42}
\bibitem{footnote1}
\bibinfo{pages}{Let us also note two related
extension of DMFT, the inhomogeneous DMFT \cite{Potthoff99} which deals with stacked macroscopically extended layers and the real 
space DMFT \cite{Snoek08} for cold atoms, 
without  environment.}
\end{thebibliography}

\end{document}